\documentclass[10pt]{iopart}
\usepackage{graphicx}
\usepackage{color}

\expandafter\let\csname equation*\endcsname\relax

\expandafter\let\csname endequation*\endcsname\relax

\usepackage[libertine]{newtxmath}
\usepackage{slashed}
\usepackage{amsfonts}

\usepackage[normalem]{ulem}

\usepackage{microtype}\usepackage[colorlinks,linkcolor=blue,anchorcolor=blue,citecolor=blue,urlcolor=blue,breaklinks=true]{hyperref}

\def\be{\begin{equation}}
\def\ee{\end{equation}}
\def\bea{\begin{eqnarray}}
\def\eea{\end{eqnarray}}

\def\S{\Sigma}

\begin{document}

\title{Dark photon effect on the rare kaon decay $K_L \rightarrow \pi^0 \nu {\bar \nu}$}

\author{Xuan-Gong Wang$^1$\footnote{To whom correspondence should be addressed.}, A W Thomas$^1$}
\address{$^1$ ARC Centre of Excellence for Dark Matter Particle Physics and CSSM, Department of Physics, University of Adelaide, Adelaide SA 5005, Australia}
\eads{\mailto{xuan-gong.wang@adelaide.edu.au}, \mailto{anthony.thomas@adelaide.edu.au}}

\begin{abstract}
We present an analysis of the effect of a dark photon on the rare kaon decay 
$K_L \rightarrow \pi^0 \nu {\bar \nu}$. 
All relevant couplings of the dark photon to the Standard Model particles are derived explicitly in terms of the dark photon mass and the mixing parameter.
We find that the dark photon yields no more than a few percent correction to the Standard Model branching ratio 
${\rm Br}(K_L \rightarrow \pi^0 \nu {\bar \nu})$ in the region of interest.
\end{abstract}


\submitto{\jpg}

\maketitle


\section{Introduction}

The dark photon is an appealing hypothesis for new physics beyond the Standard Model (SM)~\cite{Fayet:1980ad, Fayet:1980rr}.
It has emerged as a canonical portal connecting the dark matter and SM   sectors~\cite{Fabbrichesi:2020wbt, Filippi:2020kii}.
While there have been numerous experimental searches at $e^+ e^-$ and hadron colliders~\cite{Merkel:2014avp, LHCb:2019vmc, BaBar:2017tiz, Banerjee:2019pds, CMS:2019buh}, there is no direct evidence for its existence so far. 
 Instead, rather strong constraints have been derived on the mixing parameter $\epsilon$, leading to an upper limit of $\epsilon \le 10^{-3}$ in both light~\cite{BaBar:2017tiz, Banerjee:2019pds} and heavy~\cite{CMS:2019buh} mass regions, with just a few gaps.
 
 Theoretical investigations have also placed constraints on the dark photon parameters,
 by exploring its contributions to many physical processes, 
 including $g-2$ of the muon~\cite{Pospelov:2008zw, Davoudiasl:2012qa}, electroweak precision observables (EWPO)~\cite{Hook:2010tw, Curtin:2014cca}, 
 $e^- p$ deep inelastic scattering (DIS)~\cite{Kribs:2020vyk, Thomas:2021lub, Yan:2022npz,Hunt-Smith:2023sdz}, parity violating electron scattering~\cite{Thomas:2022qhj, Thomas:2022gib}, 
 and partial wave unitarity~\cite{Hosseini:2022urq}.
The dark photon can also contribute to rare kaon decays and analogous rare $B$ processes, 
which are powerful tools to test the SM~\cite{Buras:2004uu} and constrain new physics~\cite{Bryman:2005xp}.

Amongst rare decays, the processes $K^+ \rightarrow \pi^+ \nu {\bar \nu}$ and $K_L \rightarrow \pi^0 \nu {\bar \nu}$ are the so-called golden channels, as their branching ratios can be computed with high precision. The most accurate SM values for the branching fractions are~\cite{Buras:2015qea}
\begin{eqnarray}
{\rm Br}(K^+ \rightarrow \pi^+ \nu \bar{\nu}) &=& (9.11 \pm 0.72) \times 10^{-11} \ ,\nonumber\\
{\rm Br}(K_L \rightarrow \pi^0 \nu \bar{\nu}) &=& (3.00 \pm 0.31) \times 10^{-11} \ ,
\end{eqnarray}
while the latest experimental measurements from NA62~\cite{NA62:2021zjw} and  KOTO~\cite{KOTO:2020prk} set upper limits at 68\% and 90\% confidence level (CL), respectively
\begin{eqnarray}
{\rm Br}(K^+ \rightarrow \pi^+ \nu \bar{\nu})& =& ( 10.6^{+4.0}_{-3.4}|_{\rm stat} \pm 0.9_{\rm syst} )\times 10^{-11}
\ ,\nonumber\\
{\rm Br}(K_L \rightarrow \pi^0 \nu \bar{\nu})_{\rm KOTO} &<& 4.9 \times 10^{-9} \ .
\end{eqnarray}
Possible anomalies between experiments and the SM predictions have been investigated in many new physics models~\cite{Fuyuto:2014cya, Davoudiasl:2012ag, deMelo:2021ers, Egana-Ugrinovic:2019wzj, Goudzovski:2022vbt}. 

A light dark photon may be expected to yield a significant contribution to branching ratios such as these because of enhancement from the propagator, compared with the $Z$ boson. 
Moreover, for an ultralight dark photon $A_D$, it is possible that the two-body decay $K\rightarrow \pi A_D$ will occur, followed by visible or invisible decays of $A_D$~\cite{Davoudiasl:2012ag, Davoudiasl:2014kua}.
Recently, a light dark photon was introduced, either on-shell or off-shell, to improve the agreement of the branching ratios between the SM predictions and the experimental measurements for a bunch of rare $B$ decays~\cite{Datta:2022zng}, with three parameters, the mass $m_{Z_D}$, and two independent mixing parameters $(\epsilon,\epsilon_Z)$. However, the best fit results required an unrealistically large value of the mixing parameter $\delta$~\cite{Davoudiasl:2012ag}.

Here, we perform a quantitative analysis of the rare kaon decays within the dark photon framework, focusing on the $K_L \rightarrow \pi^0 \nu {\bar \nu}$ channel in order to avoid the complexity arising from the charm quark contribution in the $K^+$ channel. 
All the couplings of the dark photon to the SM particles are explicitly dependent on just two parameters, the dark photon mass, $m_{A_D}$, and the mixing parameter,  $\epsilon$.
We also explore the sensitivity of its branching ratio to the dark photon parameters.

In Sec.~\ref{sec:SM}, we briefly review the rare kaon decays in the Standard Model. 
We derive the couplings of the physical $Z$ and the dark photon in Sec.~\ref{sec:Weak-couplings} and~~\ref{sec:Dark-couplings},
 and the correction to the branching ratio in Sec.~\ref{sec:br}.
We present our numerical results in Sec.~\ref{sec:results}.
Finally, we summarize our analysis in Sec.~\ref{sec:conclusion}.

\section{Standard Model predictions}
\label{sec:SM}

The SM contributions to the decay modes $K^+ \rightarrow \pi^+ \nu \bar{\nu}$ and $K_L \rightarrow \pi^0 \nu \bar{\nu}$ 
 include the ``$Z$ penguin" and the ``box" diagrams with up, charm, and top quark exchanges.
The invariant amplitudes  can be written in the following form~\cite{Buchalla:1995vs},
\be
\label{eq:amplitude}
{\cal A} = - i \frac{G_F}{\sqrt{2}} \frac{\alpha_{em}}{2\pi \sin^2\theta_W}
\sum_{l=e, \mu, \tau} \left(V^*_{cs} V_{cd} X^l_{NL} + V^*_{ts} V_{td} X(x_t) \right) 
 \times (\bar{s} d)_{V-A} (\bar{\nu}_l \nu_l)_{V-A} \, .
\ee
The function $X^l_{NL}$ represents the charm quark contribution, which is only relevant to the $K^+ \rightarrow \pi^+ \nu \bar{\nu}$ channel and   
results from the renormalization group calculation in next-to-leading-order logarithmic approximation.

In this work, we will focus on the $K_L \rightarrow \pi^0 \nu \bar{\nu}$ decay, which only depends on the corresponding function in the top quark sector, $X(x_t) \equiv X(x_t, y_l = 0)$, by neglecting the lepton masses. 
Including next-to-leading order (NLO) Quantum Chromodynamics (QCD) corrections, it is given by~\cite{Buchalla:1995vs}
\be
\label{eq:X_Zbar}
X(x_q,y_l) = X^{(0)}(x_q,y_l) + \frac{\alpha_s}{4\pi} X^{(1)}(x_q,y_l) \, ,
\ee
where $x_q = m_q^2/m^2_W$ and $y_l = m_l^2/m_W^2$, and 
\be
\label{eq:X_Zbar}
X^{(n)}(x_q,y_l) = C^{(n)}_{\bar Z}(x_q) - 4 B^{(n)}(x_q,y_l) \, , \ \ \ (n = 0, 1) \, ,
\ee
with $C^{(n)}_{\bar Z}(x_q)$ and $B^{(n)}(x_q, y_l)$ the ``${\bar Z}$ penguin" and the ``box" contributions, respectively.

It is convenient to compute the loop functions in $^{'}$t Hooft-Feynman gauge ($\xi=1$), 
in which both the induced ${\bar s} d {\bar Z}$ vertex and the box diagram also receive contributions from the would-be Goldstone bosons ($\phi^{\pm}$)~\cite{Inami:1980fz}.
The leading order (LO) terms  are given by~\cite{Inami:1980fz, Buchalla:1990qz} (see the Appendix~\ref{sec:A1}),
\begin{subequations}
\begin{alignat}{4}
\label{eq:C0-Zbar}
C_{\bar Z}^{(0)}(x) &= \frac{x}{8} \Big[ \frac{x-6}{x-1} + \frac{3 x + 2}{(x-1)^2} \ln x \Big] \, ,\\
\label{eq:B0}
B^{(0)}(x,y) &= \frac{1}{64} \Big[ \frac{16 + x y - 8 y}{x - y} \frac{x^2 \ln x}{(1-x)^2} 
 - \frac{x y}{x - y} \left( \frac{y-4}{y-1} \right)^2 \ln y  + \frac{16 x - 7 x y}{(1-x)(1-y)} \Big] \, .
\end{alignat}
\end{subequations}
The QCD correction in the top quark sector is~\cite{Buras:2004uu}
\begin{alignat}{4}
X^{(1)}(x_t) &= - \frac{29 x_t - x_t^2 - 4 x_t^3}{3(1 - x_t)^2} 
- \frac{x_t + 9 x_t^2 - x_t^3 - x_t^4}{(1 - x_t)^3} \ln x_t  
 + \frac{8 x_t + 4 x_t^2 + x_t^3 - x_t^4}{2(1 - x_t)^3} \ln^2 x_t \nonumber\\
&  - \frac{4 x_t - x_t^3}{(1-x_t)^2} L_2(1-x_t)
 + 8 x_t \frac{\partial X^{(0)}(x_t)}{\partial x_t} \ln x_{\mu} \, , 
\end{alignat}
where $x_{\mu} = \mu_t^2 / M_W^2$, $\mu_t = {\cal O} (m_t)$ and 
\be
L_2(1-x_t) = \int_1^{x_t} dt \frac{ \ln t }{1-t} \, .
\ee
The branching ratio for $K_L \rightarrow \pi^0 \nu\bar{\nu}$ involves only the top-quark contribution and can be written as~\cite{Buchalla:1996fp, Buchalla:1995vs}
\be
{\rm Br}(K_L \rightarrow \pi^0 \nu\bar{\nu}) = \kappa_L \left( \frac{{\rm Im} \lambda_t}{\lambda^5} X(x_t) \right)^2 \, ,
\ee
where $\lambda_t = V^*_{t s} V_{t d}$ are the Cabibbo-Kobayashi-Maskawa (CKM) factors, 
and $\kappa_L$ parametrizes the hadronic matrix element~\cite{Marciano:1996wy, Mescia:2007kn},
\be
\kappa_L = (2.231 \pm 0.013) \times 10^{-10} \left( \frac{\lambda}{0.225} \right)^8\, .
\ee
%


\section{Dark photon formalism}
\label{sec:SM-DP}
The dark photon is usually introduced as an extra $U(1)$ gauge boson~\cite{Fayet:1980ad, Fayet:1980rr, Holdom:1985ag}, 
interacting with the SM particles through kinetic mixing with hypercharge~\cite{Okun:1982xi}
\bea
\mathcal{L} & \supset & {\cal L}_{\rm SM}^{\rm int} - \frac{1}{4} F_{\mu\nu} F^{\mu\nu} - \frac{1}{4} {\bar Z}_{\mu\nu} {\bar Z}^{\mu\nu} + \frac{1}{2} m^2_{\bar Z} {\bar Z}_{\mu} {\bar Z}^{\mu} \nonumber\\
&& - \frac{1}{4} F'_{\mu\nu} F'^{\mu\nu} + \frac{m^2_{A'}}{2} A'_{\mu} A'^{\mu} + \frac{\epsilon}{2 \cos\theta_W} F'_{\mu\nu} B^{\mu\nu} \, ,
\eea
where $\theta_W$ is the Weinberg angle. 
$A'$ and $\bar{Z}$ denote the unmixed version of the dark photon and the SM neutral weak boson, respectively.
$F_{\mu\nu}$ and ${\bar Z}_{\mu\nu}$ are the SM field strength tensors.

After diagonalizing the mixing term through field redefinitions, the masses of the physical $Z$ and $A_D$ are given by~\cite{Gopalakrishna:2008dv, Kribs:2020vyk}
\be
\label{eq:m_Z_AD}
M^2_{Z, A_D} = \frac{m_{\bar{Z}}^2}{2} \Big[ 1 + \epsilon_W^2 + \rho^2 
 \pm {\rm sign}(1-\rho^2) \sqrt{(1 + \epsilon_W^2 + \rho^2)^2 - 4 \rho^2} \Big] \, ,
\ee
where
\begin{eqnarray}
\epsilon_W &=& \frac{\epsilon \tan \theta_W}{\sqrt{1 - \epsilon^2/\cos^2\theta_W}}\,  ,\nonumber\\
\rho &=& \frac{m_{A'}/m_{\bar{Z}}}{\sqrt{1 - \epsilon^2/\cos^2\theta_W}} \, .
\end{eqnarray}
The $\bar{Z}-A'$ mixing angle $\alpha$ is given by
\be
\tan \alpha = \frac{1}{2\epsilon_W} \Big[ 1 - \epsilon^2_W - \rho^2 
 - {\rm sign}(1-\rho^2) \sqrt{4\epsilon_W^2 + (1 - \epsilon_W^2 - \rho^2)^2} \Big] \, .
\ee
The mass difference between the physical $Z$ and $A_D$ is always finite for non-zero $\epsilon$, $|M^2_Z - M^2_{A_D}| \ge 2 |\epsilon_W| m^2_{\bar{Z}}$. Thus there is a region of the dark photon parameter space which is inaccessible~\cite{Kribs:2020vyk}, which is known as the ``eigenmass repulsion" region on the $\epsilon - M_{A_D}$ plane.

Because of kinetic mixing, the SM weak couplings of the $Z$ boson given by~(\ref{eq:C-Zbar-fermion}) and~(\ref{eq:C-Zbar-gauge}) will be modified. 
The dark photon will also couple to the SM particles.
All the physical couplings depend on only two parameters, $M_{A_D}$ and $\epsilon$.

\subsection{Weak couplings}
\label{sec:Weak-couplings}
The couplings of the physical $Z$ to the quarks are given by~\cite{Kribs:2020vyk, Thomas:2022qhj, Gopalakrishna:2008dv}
\bea
C_{Z,q}^v &=& (\cos\alpha - \epsilon_W \sin\alpha) C_{\bar{Z}, q}^v + 2 \epsilon_W \sin\alpha \cos^2 \theta_W C_{\gamma, q}^v \, ,\nonumber\\
C_{Z,q}^a &=& (\cos\alpha - \epsilon_W \sin\alpha) C_{\bar{Z}, q}^a \, ,
\eea
where $C_{\gamma, u}^v= 2/3$ and $C_{\gamma, d}^v = - 1/3$.
Its couplings to the neutrinos will be shifted by
\bea
C_{Z,\nu_e} &=& (\cos\alpha - \epsilon_W \sin\alpha) C_{{\bar Z},\nu_e} \, .\
\eea
The SM couplings of the ${\bar Z}$ to the gauge bosons and the Goldstone bosons will also be modified as
\bea
C_{Z,WW} &=& ( \cos\alpha - \epsilon_W \sin\alpha ) C_{{\bar Z},WW} + 2 \epsilon_W \sin\alpha \cos^2\theta_W \, , \nonumber\\
C_{Z,\phi^{\pm}} &=& ( \cos\alpha - \epsilon_W \sin\alpha )\ C_{{\bar Z},\phi^{\pm}} + 2 \epsilon_W \sin\alpha \cos^2\theta_W\, , \nonumber\\
C_{Z,W\phi} &=& ( \cos\alpha - \epsilon_W \sin\alpha) C_{{\bar Z},W\phi} - 2 \epsilon_W \sin\alpha \cos^2\theta_W \ .
\eea

\subsection{Dark couplings}
\label{sec:Dark-couplings}
\noindent The dark photon interacts with the quarks ($q = u, d$) through both vector and axial-vector couplings~\cite{Kribs:2020vyk, Thomas:2022qhj, Gopalakrishna:2008dv},
\bea
C_{A_D, q}^v &=& - (\sin\alpha + \epsilon_W \cos\alpha) C_{\bar{Z}, q}^v + 2 \epsilon_W \cos\alpha \cos^2 \theta_W C_{\gamma, q}^v \, ,\nonumber\\
C_{A_D, q}^a &=& - (\sin\alpha + \epsilon_W \cos\alpha) C_{\bar{Z}, q}^a \, . 
\eea
Its interaction with the neutrinos also has $V-A$ form, with
\bea
C_{A_D,\nu_l} &=&  - (\sin\alpha + \epsilon_W \cos\alpha) C_{{\bar Z},\nu_e} \, .
\eea
We also derive its couplings to the gauge bosons and the Goldstone bosons, 
\bea
C_{A_D,WW} &=&  - ( \sin\alpha + \epsilon_W \cos\alpha ) C_{{\bar Z},WW} + 2 \epsilon_W \cos\alpha \cos^2\theta_W \, ,\nonumber\\ 
C_{A_D,\phi^{\pm}} &=& - ( \sin\alpha + \epsilon_W \cos\alpha ) C_{{\bar Z},\phi^{\pm}} + 2 \epsilon_W \cos\alpha \cos^2\theta_W \, , \nonumber\\
C_{A_D,W\phi} &=& - ( \sin\alpha + \epsilon_W \cos\alpha ) C_{{\bar Z},W\phi} - 2 \epsilon_W \cos\alpha \cos^2\theta_W \, .
\eea

\subsection{Branching ratio}
\label{sec:br}
The most general form of the matrix element $X^{(0)}(x_q,y_l)$, after restoring the $Z$-boson propagator and its coupling to neutrinos, can be written as
\be
\label{eq:X0-Zbar}
X^{(0)}_{\rm SM}(x_q,y_l) = - \frac{2 m^2_W }{\cos^2\theta_W} \frac{C_{{\bar Z}, \nu_l}}{k^2 - m^2_{\bar Z}} C^{(0)}_{\bar Z}(x_q) - 4 B^{(0)}(x_q,y_l) \, ,
\ee
where $k$ is the momentum transfer through the ${\bar Z}$ propagator.
When dark photon effects are included, the above expression is generalised to 
\be
X^{(0)}(x_q,y_l) 
= - \frac{2 m^2_W}{\cos^2\theta_W} \frac{ C_{Z, \nu_l}}{k^2 - M^2_Z} C^{(0)}_{Z}(x_q) 
- \frac{2 m^2_W}{\cos^2\theta_W} \frac{ C_{A_D, \nu_l}}{k^2 - M^2_{A_D}} C^{(0)}_{A_D}(x_q) 
- 4 B^{(0)}(x_q,y_l) \, ,
\ee
where the functions $C^{(0)}_{Z}(x_q)$ and $C^{(0)}_{A_D}(x_q)$ have the same form as $C^{(0)}_{\bar Z}(x_q)$ (see Appendix~\ref{sec:A1}), 
with the ${\bar Z}$ couplings being replaced by those for $Z$ and $A_D$, respectively.
Note that the inclusion of the dark photon does not affect the box contribution.

The dark photon effect can be characterised by a correction factor to the SM branching ratio,
\be
\frac{{\rm Br}(K_L \rightarrow \pi^0 \nu\bar{\nu})}{{\rm Br}(K_L \rightarrow \pi^0 \nu\bar{\nu})|_{\rm SM}} = 1 + R_L \, , 
\ee
which is independent of $\kappa_L$, $\lambda_t$, and $\lambda$.


\section{Numerical results}
\label{sec:results}
\begin{figure}[!h]
\includegraphics[width=0.9\columnwidth]{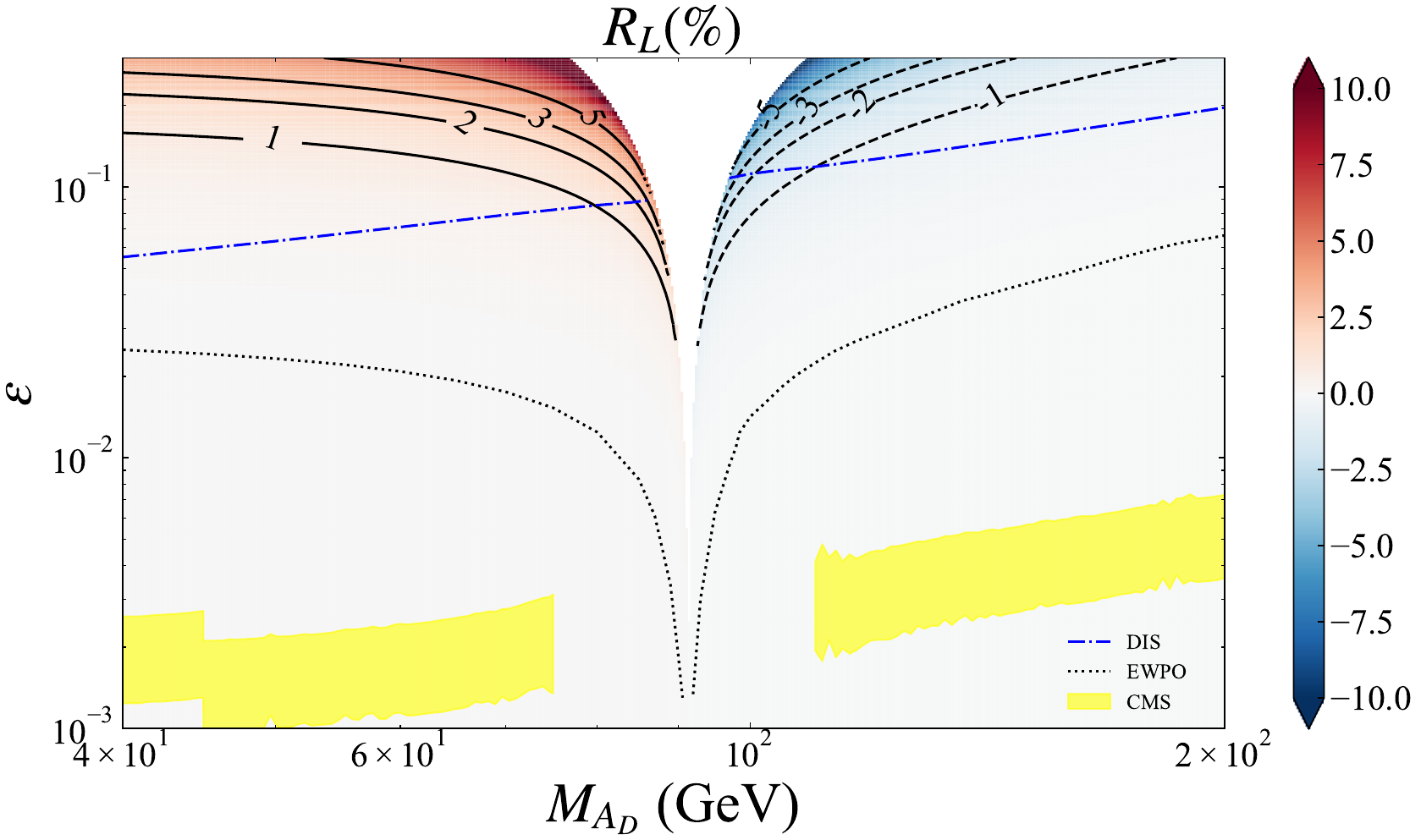}
\vspace*{-0.1cm}
\caption{ The dark photon correction to the SM value of the branching ratio ${\rm Br}(K_L \rightarrow \pi^0 \nu\bar{\nu})$.
The gap on the $\epsilon-M$ plane is not accessible due to ``eigenmass repulsion" associated with the $Z$ mass. The 95\% CL exclusion limits on $\epsilon$ from DIS and EWPO determinations are taken from Refs.~\cite{Thomas:2021lub, Thomas:2022gib} and Ref.~\cite{Curtin:2014cca}, respectively. We also show the most stringent constraint from the CMS Collaboration~\cite{CMS:2019buh}.}
\label{fig:R}
\end{figure}

In the numerical analysis, we take the parameters~\cite{Buras:2015qea}
\be
\sin^2\theta_W (M_Z) = 0.23116 \, ,\ \alpha_s(M_Z) = 0.118,\ m_t = 161\ {\rm GeV} \, .
\ee
The strongest experimental constraint on $\epsilon$ comes from the CMS Collaboration~\cite{CMS:2019buh}, leading to an upper limit of $\epsilon \sim 10^{-3}$, while the region of parameter space with $M_Z < M_{A_D} < 110\ {\rm GeV}$ is unconstrained as the $Z$ boson dominates $\mu^+\mu^-$ production there. The current limit in connection with electroweak precison observables (EWPOs)~\cite{Hook:2010tw, Curtin:2014cca} leads to $\epsilon < 0.06$ for dark photon mass up to 200 GeV, which becomes much stronger when $M_{A_D}$ gets close to $M_Z$. The recent $e^- p$ DIS analysis~\cite{Thomas:2021lub, Thomas:2022gib} placed relatively weak constraint on $\epsilon$, which will go above 0.1 when $M_{A_D} > M_Z$. Therefore, the dark photon parameters in the region $\epsilon \le 0.2$ in the $(\epsilon, M_{A_D})$ plane is of most interest, as this region has not been fully excluded by the existing constraints~\cite{Thomas:2022qhj}.
The corrections to the SM prediction of the branching ratio ${\rm Br}(K_L \rightarrow \pi^0 \nu\bar{\nu})$ are shown in figure~\ref{fig:R} as a percentage.
Surprisingly, the sensitivity of the correction factor $R_L$ to the dark photon parameters is quite similar to that of $C_{1q}$ at low scale in parity violating electron scattering~\cite{Thomas:2022qhj}. $R_L$ is at most several percent, corresponding to the case where the dark photon parameters approach the ``eigenmass repulsion" region.

It might have been expected that a light dark photon with mass at sub-GeV region could lead to a large correction.
However, the dark couplings become negligibly small in that region because they scale as $M_{A_D}/M_{Z}$, eliminating the enhancement associated with the propagator.
It would also be interesting to apply the current framework to two-body decay $K \rightarrow \pi A_D$ for ultralight dark photon.

%


\section{Conclusion}
\label{sec:conclusion}
We have presented a systematic calculation of the dark photon contribution to the rare kaon decays, focusing on the channel $K_L \rightarrow \pi^0 \nu\bar{\nu}$.
We explicitly derived the coupling constants of the dark photon to the Standard Model fermions, the gauge bosons, and the would-be Goldstone bosons in the $^{'}$t Hooft-Feynman gauge.

In contrast with naive expectations, the branching ratio of $K_L \rightarrow \pi^0 \nu\bar{\nu}$ deviates from the Standard Model prediction by at most a few percent. This deviation would be at its largest were a dark photon to exist with parameters close to the ``eigenmass repulsion" region. Unfortunately this effect is too small to be observed in the near future, given the experimental accuracy anticipated in the next few years~\cite{Moulson:2019ifj}. However, the advances in precision of such important tests of the Standard Model improve inexorably.

On the other hand, we expect that the dark photon would have a similar effect on the branching ratio of the charged channel $K^+ \rightarrow \pi^+ \nu \bar{\nu}$, while a quantitative analysis requires inclusion of the charm quark sector which adds 30\% to the total branching ratio. 
Experiments at the Brookhaven National Laboratory~\cite{BNL-E949:2009dza} has measured ${\rm Br}(K^+ \rightarrow \pi^+ \nu \bar{\nu}) = ( 17.3^{+11.5}_{-10.5} )\times 10^{-11}$.
As mentioned earlier, a more precise result was reported by the NA62 experiment~\cite{NA62:2021zjw}, ${\rm Br}(K^+ \rightarrow \pi^+ \nu \bar{\nu}) = ( 10.6^{+4.0}_{-3.4}|_{\rm stat} \pm 0.9_{\rm syst} )\times 10^{-11}$ at 68\% CL.
The proposed HIKE experiment is expected to reach a branching ratio measurement with ${\cal O}(5\%)$ precision~\cite{HIKE:2022qra}, making the charged channel more promising as a probe of the dark photon.

Finally, it is worth noting that the dark photon may serve as a promising portal connecting dark matter and ordinary particles. One can also extend the current analysis by introducing the dark photon couplings to dark matter particles. The potential anomalies of rare kaon decays may then be applied to constrain these extra invisible modes, in addition to the neutrino final states.


\ack{
This work was supported by the University of Adelaide and the Australian Government through the Australian Research Council Centre of Excellence for Dark Matter Particle Physics (CDMPP, CE200100008).
}

\appendix

\section{Induced $\bar{s} d {\bar Z}$ vertex}
\label{sec:A1}

\begin{figure}[!t]
\includegraphics[width=\columnwidth]{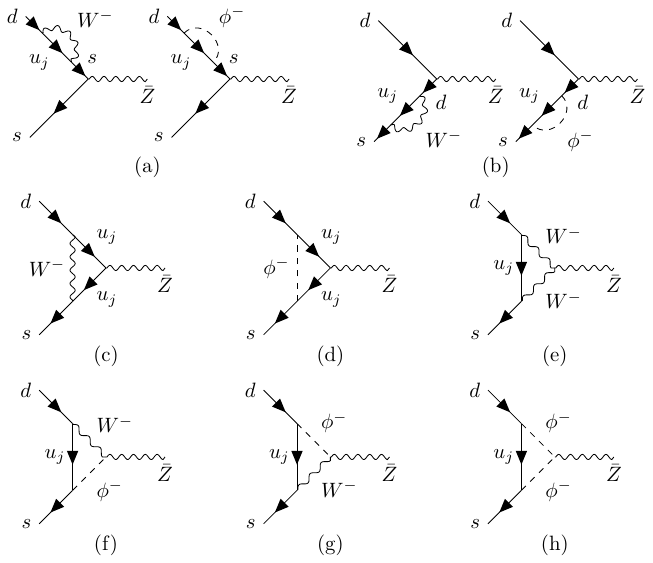}
\vspace*{-0.1cm}
\caption{ The penguin diagrams contributing to the induced ${\bar s}d {\bar Z}$ vertex.}
\label{fig:Zbar-penguin}
\end{figure}

The penguin diagrams of the induced ${\bar s}d {\bar Z}$ vertex are shown in figure~\ref{fig:Zbar-penguin}, where $u_j$ denotes the up-type quark of the $j$-th generation.
Here we rewrite the individual contributions given by~\cite{Inami:1980fz} explicitly in terms of the weak couplings as
\begin{alignat}{4}
\label{eq:Gamma-Zbar}
\Gamma^{(a+b)}_{\bar Z} &=
 \frac{1}{4} (C_{{\bar Z}, d}^v + C_{{\bar Z}, d}^a ) \Big[ \frac{x_j^2}{(x_j - 1)^2} \ln x_j - \frac{x_j}{x_j-1} - x_j f_1(x_j) \Big] 
  - (x_j \rightarrow x_1) \, ,\nonumber\\
\Gamma^{(c)}_{\bar Z} &=
 \frac{1}{4} C_{{\bar Z}, u}^v  \Big[ \frac{1}{x_j - 1} - \frac{x_j^2}{(x_j - 1 )^2 } \ln x_j \Big] 
 +  \frac{1}{4} C_{{\bar Z}, u}^a  \Big[ - \frac{3x_j}{x_j-1} - \frac{x_j (x_j - 4)}{(x_j - 1 )^2 } \ln x_j \Big] - (x_j \rightarrow x_1) \, ,\nonumber\\
 \Gamma^{(d)}_{\bar Z} &=
 \frac{1}{4} ( C_{{\bar Z}, u}^v - C_{{\bar Z}, u}^a ) (1 - \frac{2}{n}) x_j f_2(x_j) \nonumber\\
& - \frac{1}{8} ( C_{{\bar Z}, u}^v + 3 C_{{\bar Z}, u}^a ) \Big[ \frac{x_j^2}{(x_j - 1)^2} \ln x_j - \frac{x_j^2}{x_j - 1} \Big] 
 - (x_j \rightarrow x_1) \, , \nonumber\\
 \Gamma^{(e)}_{\bar Z} &=
 \frac{3}{4} C_{{\bar Z},WW} \Big[ \frac{x_j^2}{(x_j - 1)^2} \ln x_j - \frac{1}{x_j-1} \Big] - (x_j \rightarrow x_1) \, , \nonumber\\
 \Gamma^{(f+g)}_{\bar Z} &=
 \frac{1}{2} C_{{\bar Z},W\phi} \Big[ \frac{x_j^2}{(x_j - 1)^2} \ln x_j - \frac{x_j}{x_j-1} \Big] - (x_j \rightarrow x_1) \, ,\nonumber\\
\Gamma^{(h)}_{\bar Z} &= 
\frac{1}{2} C_{{\bar Z},\phi^{\pm}} \left\{ \frac{1}{4} \Big[ \frac{x_j^2}{(x_j - 1)^2} \ln x_j - \frac{x_j}{x_j-1} \Big] - \frac{1}{n} x_j f_2(x_j) \right\}
 - (x_j \rightarrow x_1) \ ,
\end{alignat}
where
\begin{alignat}{4}
f_1(x) &= - \frac{1}{n-4} + \frac{1}{2} \Big[ - \gamma_E + \ln 4\pi - \ln m_W^2 \Big] 
 + \frac{3}{4} - \frac{1}{2} \Big[ \frac{x^2}{(x-1)^2} \ln x - \frac{1}{x-1} \Big]\, ,\nonumber\\
f_2(x) &= - \frac{2}{n-4} + \Big[ - \gamma_E + \ln 4\pi - \ln m_W^2 \Big]  + 1 - \frac{x}{x-1} \ln x\, .
\end{alignat}
Note that in~(\ref{eq:Gamma-Zbar}) the light quark ($u_1$) contribution has been rearranged into the other two heavy quark contributions, by use of the unitarity of the CKM matrix. That is why the total amplitude in~(\ref{eq:amplitude}) only involves the sum of charm and top quark contributions.

The penguin functions $C^{(0)}_{\bar Z}(x_j)$ for $j=2,3$ are
\bea
C^{(0)}_{\bar Z}(x_j) = \frac{1}{2} \Gamma_{\bar Z} &=& \frac{1}{2} \sum_{i=a}^{h} \Gamma^{(i)}_{\bar Z} \, ,
\eea
giving rise to~(\ref{eq:C0-Zbar}).

\section{SM couplings}
\label{sec:A2}
In the SM, the weak couplings of ${\bar Z}$ to the up-type ($u$) and down-type ($d$) quarks and to the neutrinos ($\nu_l$) are 
\bea
\label{eq:C-Zbar-fermion}
C^v_{\bar{Z}, u} &=& \frac{1}{2} - \frac{4}{3}\sin^2\theta_W \, ,\ C^a_{\bar{Z}, u} = \frac{1}{2} \, , \nonumber\\
C^v_{\bar{Z}, d} &=& - \frac{1}{2} + \frac{2}{3}\sin^2\theta_W \, ,\ C^a_{\bar{Z}, d} = - \frac{1}{2} \, \nonumber\\
C_{{\bar Z},\nu_l} &=& \frac{1}{2} \ ,
\eea
where the superscripts $v$ and $a$ denote the vector and the axial-vector components, respectively.

The gauge couplings in~(\ref{eq:Gamma-Zbar}) are given by
\bea
\label{eq:C-Zbar-gauge}
C_{{\bar Z},WW} &=& 2 \cos^2 \theta_W \, ,\nonumber\\
C_{{\bar Z},\phi^{\pm}} &=& 1 - 2 \sin^2 \theta_W \, ,\nonumber\\
C_{{\bar Z},W\phi} &=& 2 \sin^2\theta_W \, .
\eea


\end{document}